\documentclass[twocolumn,showpacs,preprintnumbers,aps]{revtex4}
\usepackage{amsmath}
\usepackage{dcolumn}
\usepackage{bm}
\usepackage{graphicx}
\usepackage{amsfonts}
\usepackage{amssymb}
\usepackage{bbm}
\usepackage{float}
\usepackage[dvipdfm]{hyperref}

\begin{document}

\title{Non-Markovian effect on remote state preparation}
\author{Zhen-Yu Xu $^{1}$}
\email{zhenyuxu@suda.edu.cn}
\author{Chen Liu $^{1}$}
\author{Shunlong Luo $^{2}$}
\email{luosl@amt.ac.cn}
\author{Shiqun Zhu $^{1}$}
\email{szhu@suda.edu.cn}
\affiliation{$^{1}$College of Physics, Optoelectronics and Energy, Soochow University, Suzhou
215006, China\\
$^{2}$Academy of Mathematics and Systems Science, Chinese Academy of
Sciences, Beijing 100190, China}

\begin{abstract}
Memory effect of non-Markovian dynamics in open quantum systems is often believed
to be beneficial for quantum information processing. In this work, we employ
an experimentally controllable two-photon open system, with one photon
experiencing a dephasing environment and the other being free from noise, to
show that non-Markovian effect may also have a negative impact on quantum
tasks such as remote state preparation: For a certain period of controlled
time interval, stronger non-Markovian effect yields lower fidelity of remote
state preparation, as opposed to the common wisdom that more information
leads to better performance. As a comparison, a positive non-Markovian effect on the RSP fidelity with another typical non-Markovian noise
is analyzed. Consequently, the observed dual character of
non-Markovian effect will be of great importance in the field of open
systems engineering.
\end{abstract}

\maketitle


\section{Introduction}

Non-Markovianity is a ubiquitous and remarkable feature in open quantum
systems with structured reservoirs or strong system-environment couplings
\cite{Book-Open,Book-Open2}. Due to the recent development of experimental
techniques, non-Markovian effect has been observed and engineered in a
variety of systems, such as high-Q cavities \cite{Cavity}, linear photon
systems \cite{photon1,photon2,photon3}, ultracold neutral plasma \cite%
{Plasma}, nuclear magnetic resonance systems \cite{NMR}, solid state systems
\cite{Solid1,Solid2}, and quantum biology \cite{QB}. Several theoretical
methods have been developed to describe non-Markovian dynamics and to
construct phenomenological descriptions \cite%
{Book-Open,Book-Open2,nonM-review1,nonM-review2}.

It is found that non-Markovian effect usually plays a positive role in many
quantum tasks such as entanglement preservation \cite{QCP}, quantum key
distribution \cite{QCommun}, quantum speedup\cite{speedup1,speedup2}, and
quantum metrology \cite{metrology1,metrology2}. The reason lies in the fact
that non-Markovian environment could be taken as a memory device which
partially records the information of the open systems and then returns it
back later. How to control the non-Markovian behavior of open systems is
important in both theory and experiments, and several non-Markovian control
experiments have been realized \cite{nonM-C-photon,nonM-C-atom,nonM-C-T}.

However, a basic problem in open systems has been largely overlooked: Does
non-Markovian effect always play a positive role in quantum tasks? This
issue is significant not only for the understanding of non-Markovian
dynamics, but also for the engineering of open systems. In this work, by
considering a controllable photonic open system, we show that non-Markovian
effect is quite a mixed blessing, which may have negative effect on certain
fundamental quantum tasks. We illustrate this phenomenon by an instance of
non-Markovian negative effect on the fidelity of remote state preparation
(RSP), in terms of several non-Markovianity measures. As a comparison, a
positive effect on the RSP fidelity with another typical non-Markovian noise
is analyzed, showing the dual character of non-Markovian effect.

\section{Physical model}

The controllable physical model of our open system is the polarization
freedom $ab$ of an entangled photon pair $AB=aa^{\prime }:bb^{\prime }$
created in spontaneous parametric down-conversion, with the frequency
freedom $a^{\prime }b^{\prime }$ serving as the environment. The
polarization system $a$ of photon $A$ will experience a dephasing
interaction with environment $a^{\prime }$, which is realized by a quartz
plate, and photon $B$ is free from noise. The dephasing of photon $A$ in a
quartz plate is caused by local interaction between the polarization system $%
a$ (open system pertaining to $A$) and the frequency freedom $a^{\prime }$
(environment pertaining to $A$) with the interaction Hamiltonian ($\hbar =1$%
) \cite{Hamiltonian}
\begin{equation}
H^{aa^{\prime }}=-\left( n_{V}\left\vert V\right\rangle _{a}\left\langle
V\right\vert +n_{H}\left\vert H\right\rangle _{a}\left\langle H\right\vert
\right) \otimes \int \omega \left\vert \omega \right\rangle _{a^{\prime
}}\left\langle \omega \right\vert d\omega ,
\end{equation}%
where $|H\rangle _{a}$ and $|V\rangle _{a}$ are the polarization, and $%
|\omega \rangle _{a^{\prime }}$ is the frequency, states of photon $A$, $%
n_{V}$ and $n_{H}$ are the refraction indexes. The interaction only occurs
in the quartz plate, and the total controllable Hamiltonian is
\begin{equation}
H_{\mathrm{c}}^{aa^{\prime }bb^{\prime }}(t)=\mu (t)H^{aa^{\prime }}\otimes
\mathbf{1}^{bb^{\prime }},
\end{equation}%
where $\mu \left( t\right) =1$ if $t\in \left[ t_{0},t_{c}\right] $ and $\mu
\left( t\right) =0$ otherwise, $t_{0(c)}$ denotes the time of photon $A$
entering (leaving) the quartz plate (i.e., the control time is related to
the thickness of the quartz plate), and $\mathbf{1}^{bb^{\prime }}=\mathbf{1}%
^{b}\otimes \mathbf{1}^{b^{\prime }}$ is the identity operator for photon $B$
with $\mathbf{1}^{b}$ and $\mathbf{1}^{b^{\prime }}$ the identity operators
for its polarization and frequency freedom, respectively. For simplicity, we
take $t_{0}=0$ hereafter.

Consider an initial two-photon state $\rho ^{ab}\otimes \rho ^{a^{\prime
}b^{\prime }}$, where $\rho ^{ab}$ is the polarization state representing
the open system, and $\rho ^{a^{\prime }b^{\prime }}=\int d\omega d\omega
^{^{\prime }}f(\omega )f^{\ast }(\omega ^{^{\prime }})\left\vert \omega
\right\rangle _{a^{\prime }}\left\langle \omega ^{^{\prime }}\right\vert
\otimes \rho ^{b^{\prime }}$ is the environmental state with $\left\vert
f(\omega )\right\vert ^{2}$ the probability density of finding photon $A$
with frequency $\omega $, and $\rho ^{b^{\prime }}$ any state of the
frequency freedom of photon $B$. The frequency probability density $%
\left\vert f(\omega )\right\vert ^{2},$ as well as the time $t_{c}$, are
under our control. The two-photon polarization state at time $t$ can be
expressed as
\begin{equation}
\rho ^{ab}(t)=\Lambda _{t}\rho ^{ab}:=\mathrm{tr}_{a^{\prime }b^{\prime
}}\left\{ U(t)(\rho ^{ab}\otimes \rho ^{a^{\prime }b^{\prime
}})U(t)^{\dagger }\right\}  \label{DD}
\end{equation}%
with $\Lambda =\{\Lambda _{t}\}$ our quantum dynamics, $U(t)=e^{-i%
\int_{0}^{t}H_{\mathrm{c}}^{aa^{\prime }bb^{\prime }}\left( t^{\prime
}\right) dt^{\prime }}.$ For $t\in \lbrack 0,t_{c}],$ we have $%
t=\int_{0}^{t}\mu \left( t^{\prime }\right) dt^{\prime }$ and
\begin{equation}
\rho ^{ab}(t)=\left(
\begin{array}{cccc}
\rho _{11} & \rho _{12} & \rho _{13}\kappa (t) & \rho _{14}\kappa (t) \\
\rho _{21} & \rho _{22} & \rho _{23}\kappa (t) & \rho _{24}\kappa (t) \\
\rho _{31}\kappa ^{\ast }(t) & \rho _{32}\kappa ^{\ast }(t) & \rho _{33} &
\rho _{34} \\
\rho _{41}\kappa ^{\ast }(t) & \rho _{42}\kappa ^{\ast }(t) & \rho _{43} &
\rho _{44}%
\end{array}%
\right) ,  \label{map}
\end{equation}%
where $\rho ^{ab}=(\rho _{ij})$ is the initial open system state, and $%
\kappa (t)=\int \left\vert f(\omega )\right\vert ^{2}e^{i\omega
(n_{V}-n_{H})t}d\omega .$ We consider a controllable two-peaked Gaussian
frequency distribution \cite{photon1,nonM-C-photon}
\begin{equation}
\left\vert f(\omega )\right\vert ^{2}=\frac{\cos ^{2}\theta }{\sqrt{2\pi }%
\sigma }e^{-\frac{(\omega -\omega _{1})^{2}}{2\sigma ^{2}}}+\frac{\sin
^{2}\theta }{\sqrt{2\pi }\sigma }e^{-\frac{(\omega -\omega _{2})^{2}}{%
2\sigma ^{2}}}
\end{equation}%
for the environment $a^{\prime },$ with $\theta \in \lbrack 0,\pi /2]$
controlling the relative weight of the two peaks, which are centered at $%
\omega _{1}$ and $\omega _{2}$, respectively, with the same width $\sigma .$
Then the dephasing rate takes the form
\begin{equation}
\left\vert \kappa (t)\right\vert =e^{-\frac{\sigma ^{2}\tau ^{2}}{2}}\sqrt{%
1-\sin ^{2}2\theta \cdot \sin ^{2}\frac{\Delta \omega \tau }{2}}  \label{kk}
\end{equation}%
with $\tau =(n_{V}-n_{H})t$ and$\ \Delta \omega =\omega _{2}-\omega _{1}.$

\begin{figure*}[tbp]
\centering
\includegraphics[width=10cm]{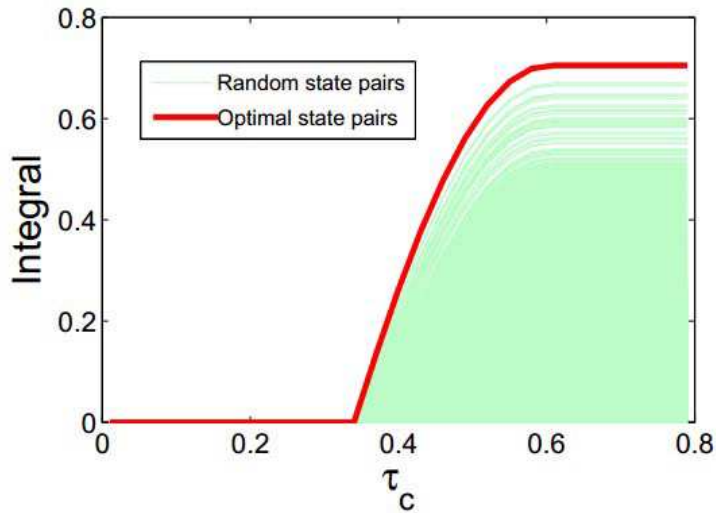}
\caption{(Color online) The integrals $\protect\int_{g>0}g(t)dt$ in the
definition of non-Markovianity, Eq. (\protect\ref{BLP}), versus control time
$\protect\tau _{c}=(n_{V}-n_{H})t_{c}$ of a two-photon system with parameter
$\protect\theta =\protect\pi /4$. The red and green curves represent the
integrals with the state pairs of Eq. (\protect\ref{optimal}) and other
10000 randomly generated pairs, respectively.}
\label{fig:wide}
\end{figure*}

To characterize the non-Markovian effect of above model, we first employ the
information-flow method \cite{BLP}, in which the non-Markovianity of $%
\Lambda =\{\Lambda _{t}\}_{t\in \lbrack 0,t_{c}]}$ is defined as the total
amount of information flowing back from the environment. The information is
quantified by the trace distance $D(\Lambda _{t}\rho _{1}^{ab},\Lambda
_{t}\rho _{2}^{ab})=(1/2)$tr$|\Lambda _{t}\rho _{1}^{ab}-\Lambda _{t}\rho
_{2}^{ab}|$ of a pair of evolved quantum states, which describes the
distinguishability between them \cite{book QIC}. The direction of
information flow depends on the gradient $g(t)=\partial _{t}D(\Lambda
_{t}\rho _{1}^{ab},\Lambda _{t}\rho _{2}^{ab})$, with positive gradient
indicating information flowing back to the system. The non-Markovianity of $%
\Lambda =\{\Lambda _{t}\}_{t\in \lbrack 0,t_{c}]}$ is quantified as \cite%
{BLP}
\begin{equation}
\mathcal{N}(\Lambda ):=\underset{\rho _{1}^{ab},\rho _{2}^{ab}}{\max }%
\int_{g>0}g(t)dt.  \label{BLP}
\end{equation}%
The integral is over $[0,t_{c}]\bigcap \{t:g(t)>0\},$ and the maximization
is over all initial state pairs $(\rho _{1}^{ab},\rho _{2}^{ab}).$ There
exists no general analytical method of finding the optimal initial state
pairs $(\rho _{1}^{ab},\rho _{2}^{ab})$ in multi-qubit cases \cite{optimal},
therefore, numerical calculations will be performed instead.

In Fig. 1, the integrals $\int_{g>0}g(t)dt$ in the definition of $\mathcal{N}%
(\Lambda )$ of 10000 random initial state pairs $(\rho _{1}^{ab},\rho
_{2}^{ab})$ (green curves) are generated with $\theta =\pi /4$. Clearly all
these pairs yield smaller values than that of the state pairs (red curve)%
\begin{equation}
\rho _{1,2}^{ab}=\left\vert \zeta ^{\pm }\right\rangle \left\langle \zeta
^{\pm }\right\vert \text{ or }\left\vert \eta ^{\pm }\right\rangle
\left\langle \eta ^{\pm }\right\vert  \label{optimal}
\end{equation}%
with
\begin{eqnarray}
\left\vert \zeta ^{\pm }\right\rangle &=&\sqrt{\alpha }\left\vert \phi ^{\pm
}\right\rangle +\sqrt{1-\alpha }\left\vert \varphi ^{\pm }\right\rangle , \\
\left\vert \eta ^{\pm }\right\rangle &=&\sqrt{\alpha }\left\vert \phi ^{\pm
}\right\rangle +\sqrt{1-\alpha }\left\vert \varphi ^{\mp }\right\rangle ,
\notag
\end{eqnarray}%
where $\alpha \in \left[ 0,1\right] ,$ $\left\vert \phi ^{\pm }\right\rangle
=\left( \left\vert HH\right\rangle \pm e^{i\theta }\left\vert
VV\right\rangle \right) /\sqrt{2},$ $\left\vert \varphi ^{\pm }\right\rangle
=$ $\left( \left\vert HV\right\rangle \pm e^{i\theta }\left\vert
VH\right\rangle \right) /\sqrt{2},$ $\theta \in \left[ 0,2\pi \right] .$ One
may check that the optimal state pairs yield the same trace distance $%
D(t)=\left\vert \kappa (t)\right\vert ,$ with $D(t)$ denoting the optimal
trace distance of non-Markovianity.

Obviously, the non-Markovianity $\mathcal{N}(\Lambda )$ depends on the
control time $t_{c}$ which characterizes the dephasing duration. For $\tau
_{c}=(n_{V}-n_{H})t_{c}\in \lbrack \pi /\Delta \omega ,2\pi /\Delta \omega
], $ we have
\begin{equation}
\mathcal{N}(\Lambda )=\left\vert \kappa (\tau _{c})\right\vert -\delta \quad
\mathrm{with}\quad \delta =\left\vert \cos 2\theta \right\vert e^{-\frac{1}{2%
}\left( \frac{\pi \sigma }{\Delta \omega }\right) ^{2}}.
\end{equation}%
\ \ \

\begin{figure*}[tbp]
\centering
\includegraphics[width=10cm]{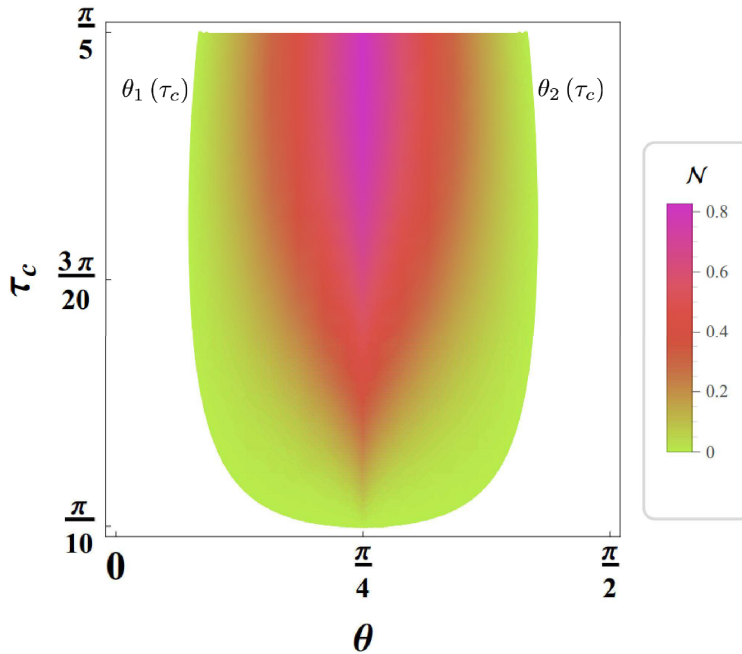}
\caption{(Color online) Non-Markovianity versus the control time $\protect%
\tau _{c}\in \lbrack \protect\pi /\Delta \protect\omega ,2\protect\pi %
/\Delta \protect\omega ]$ and the frequency control parameter $\protect%
\theta \in \lbrack 0,\protect\pi /2]$ of a two-photon system with photon $A$
suffering from dephasing noise (realized by a quartz plate with $\Delta
\protect\omega =10$ and $\protect\sigma =1$) and photon $B$ being free from
noise. $\protect\theta _{1}(\protect\tau _{c})$ denotes the sudden
transition point from Markovian to non-Markovian regime, and $\protect\theta %
_{2}(\protect\tau _{c})$ the transition point from non-Markovian to
Markovian regime.}
\end{figure*}

The non-Markovianity is depicted in Fig. 2. Clearly, for a fixed control
time $\tau _{c}$, the non-Markovianity can be adjusted by the parameter $%
\theta $. There exist two critical points of sudden transition between
Markovian and non-Markovian regime for the open system, which take the form%
\begin{equation}
\theta _{1}\left( \tau _{c}\right) =\arctan \sqrt{p-q},\quad \theta
_{2}\left( \tau _{c}\right) =\arctan \sqrt{p+q},
\end{equation}%
where
\begin{eqnarray}
p &=&\frac{u+v\cos (\Delta \omega \tau _{c})}{u-v},\quad u:=e^{\sigma
^{2}\tau _{c}^{2}},\ v:=e^{\left( \pi \sigma /\Delta \omega \right) ^{2}}
\notag \\
q &=&\frac{\sqrt{2uv(1+\cos (\Delta \omega \tau _{c}))-v^{2}\sin ^{2}(\Delta
\omega \tau _{c})}}{u-v}.
\end{eqnarray}%
This transition is similar to the one photon case observed in laboratory
\cite{nonM-C-photon}.

\section{Non-Markovian negative effect}

The above model reveals an interesting phenomenon: From Fig. 2, for $\tau
_{c}\in (\pi /\Delta \omega ,2\pi /\Delta \omega )$, we see that when $%
\theta $ varies from $\theta _{1}(\tau _{c})$ to $\pi /4$, the
non-Markovianity increases, but the trace distance $D(t)=|\kappa (t)|$
decreases in view of Eq. (\ref{kk}) for any fixed $t$, i.e., the stronger
the non-Markovian effect, the smaller the trace distance for a pair of
evolved quantum states. In other words, here the non-Markovianity renders
two quantum states less distinguishable, which turns out to be the very
reverse of the folklore intuition of non-Markovian effect making quantum
states more distinguishable due to the back flow of information.
Consequently, the illustrated non-Markovianity becomes a negative effect in
trace distance related quantum tasks. However, we note that for $\tau
_{c}=2\pi /\Delta \omega ,$ the trace distance $D(t)=\left\vert \kappa
(t_{c})\right\vert =e^{-2\left( \frac{\pi \sigma }{\Delta \omega }\right)
^{2}}$ is independent of the parameter $\theta ,$ and thus trace distance
related quantum tasks will be immune from the non-Markovian effect in this
instance.

We illustrate the implication of the above phenomenon on the RSP fidelity
\cite{RSP,RSP2}. RSP is a variation of quantum teleportation requiring only
one cbit (classical bit) and a single qubit measurement. In RSP, Alice and
Bob initially share a maximally entangled state. A qubit $\left\vert \psi
\right\rangle $ to be sent from Alice to Bob is initially known to Alice but
unknown to Bob. Alice performs a measurement on her party of the shared
entangled state along the basis \{$\left\vert \psi \right\rangle ,\left\vert
\psi _{\perp }\right\rangle $\}. The outcome of Alice's measurement is sent
to Bob with one cbit communication through a classical channel, Bob then
performs a unitary operation on his own party of the shared state according
to the cbit received. The state $\left\vert \psi \right\rangle $ can be
reconstructed with 100\% by Bob. However, if the shared state is not
maximally entangled, the success probability will decrease.

\begin{figure*}[tbp]
\centering
\includegraphics[width=12cm]{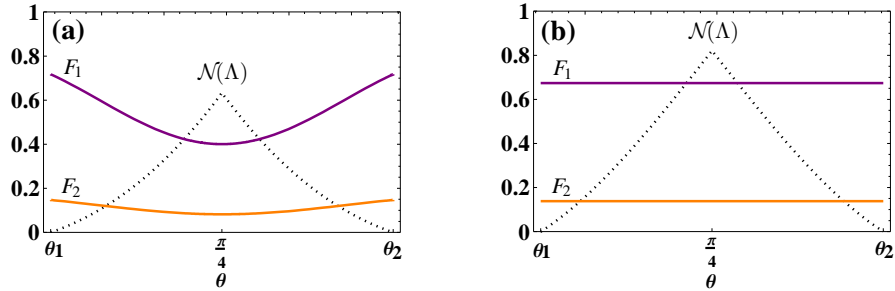}
\caption{(Color online) The dotted curve depicts the non-Markovianity with
control time $\protect\tau _{c}=3\protect\pi /(2\Delta \protect\omega )$ in
(a) and $\protect\tau _{c}=2\protect\pi /\Delta \protect\omega $ in (b)
versus the parameter $\protect\theta \in \lbrack \protect\theta _{1}(\protect%
\tau _{c}),\protect\theta _{2}(\protect\tau _{c})]$ (non-Markovian regime).
The purple (dark) and orange (light) curves represent the RSP fidelity $%
F_{1} $ and $F_{2}$ with initial Bell state $(c_{1},c_{2},c_{3}$)=(1,-1,1)
and a mixed entangled state $(c_{1},c_{2},c_{3}$)=(-0.5,0.4,0.8),
respectively.}
\end{figure*}

Now suppose that for RSP, the shared entangled state between Alice and Bob
is the decohered state $\rho ^{ab}(t)=\Lambda _{t}\rho ^{ab}$ of the
Bell-diagonal state \cite{Bell}
\begin{equation}
\rho ^{ab}=\frac{1}{4}\Big (\mathbf{1}^{a}\otimes \mathbf{1}%
^{b}+\sum_{j=1}^{3}c_{_{j}}\sigma _{j}^{a}\otimes \sigma _{j}^{b}\Big ),
\label{BD}
\end{equation}%
under the quantum dynamics $\Lambda =\{\Lambda _{t}\}$ described by Eq. (\ref%
{DD}). Here $\sigma _{j}$ are the Pauli matrices and $c_{j}$ are real
constants. Then the decohered state takes the form
\begin{equation}
\rho ^{ab}(t)=\frac{1}{4}\Big (\mathbf{1}^{a}\otimes \mathbf{1}%
^{b}+\sum_{j,j^{\prime }=1}^{3}c_{_{jj^{\prime }}}(t)\sigma _{j}^{a}\otimes
\sigma _{j^{\prime }}^{b}\Big ),
\end{equation}%
with the correlation matrix
\begin{equation}
C=(c_{jj^{\prime }})=\left(
\begin{array}{ccc}
c_{1}\mathrm{Re}\kappa (t) & c_{2}\mathrm{Im}\kappa (t) & 0 \\
-c_{1}\mathrm{Im}\kappa (t) & c_{2}\mathrm{Re}\kappa (t) & 0 \\
0 & 0 & c_{3}%
\end{array}%
\right) .
\end{equation}%
The RSP fidelity can be calculated as $F=F(C^{T}C)$ \cite{RSP-exp}$,$ which
is equal to the mean value of the two lowermost eigenvalues of the matrix $%
C^{T}C$ $($the superscript $T$ represents transposition$)$. It is convenient
to check that three eigenvalues of $C^{T}C$ are $c_{1}^{2}|\kappa (t)|^{2}$,
$c_{2}^{2}|\kappa (t)|^{2}$ and $c_{3}^{2}.$ When $c_{3}\geq c_{1},c_{2},$
we have $F=\frac{c_{1}^{2}+c_{2}^{2}}{2}|\kappa (t)|^{2}.$ In Fig. 3(a), we
plot the dependence of the RSP fidelity on non-Markovianity with a specific
control time $\tau _{c}=3\pi /(2\Delta \omega )$ for two initially entangled
states $\rho ^{ab}$ with parameters $%
(c_{1},c_{2},c_{3})=(1,-1,1),(-0.5,0.4,0.8),$ respectively. An interesting
phenomenon can be seen in Fig. 3(a): In the non-Markovian region, i.e., $%
\theta \in \lbrack \theta _{1}(\tau _{c}),\theta _{2}(\tau _{c})],$ stronger
non-Markovian effect will induce lower RSP fidelity, which implies that the
information flowed back from the environment is not always beneficial. We
also note that for some specific control time, e.g., $\tau _{c}=2\pi /\Delta
\omega ,$ the RSP fidelity is given by $F=\frac{c_{1}^{2}+c_{2}^{2}}{2}%
e^{-\left( \frac{2\pi \sigma }{\Delta \omega }\right) ^{2}},$ which is
totally uncorrelated with the parameter $\theta $, as shown in Fig. 3(b).
The reason is that, in this situation, no matter how strong non-Markovianity
is, the trace distance evolves to the same value, and the RSP fidelity
therefore keeps unchanged.

\section{Other measures for non-Markovianity}

Apart from the information-flow based non-Markovianity measure, there are
many other measures quantifying non-Markovianity \cite%
{nonM-review1,nonM-review2,RHP,Lu,LFS}, and they do not coincide in general
\cite{Compare1,Compare2,Compare3,JM}. The question arises: Does the negative
non-Markovian effect arise solely due to information-flow based
non-Markovianity measure, or is it a rather general feature which also
survives for other non-Markovianity measures? To answer this question, we
further consider three other popular measures for non-Markovianity.

In Ref. \cite{RHP}, a divisibility based non-Markovianity measure is defined
as
\begin{equation}
\mathcal{N}_{D}(\Lambda ):=\int_{0}^{\infty }h(t)dt
\end{equation}%
with $h(t)=\underset{\epsilon \rightarrow 0}{\lim }\frac{\mathrm{tr}%
|(\Lambda _{t+\epsilon ,t}\otimes \mathbf{I})\rho ^{ss^{\prime }}|-1}{%
\epsilon }$. Here $\Lambda _{t+\epsilon ,t}$ is defined via $\Lambda
_{t+\epsilon }=\Lambda _{t+\epsilon ,t}\Lambda _{t},\ \epsilon \geq 0,$ $%
\rho ^{ss^{\prime }}=\left\vert \Psi \right\rangle \left\langle \Psi
\right\vert $ with $\left\vert \Psi \right\rangle =\frac{1}{\sqrt{d}}%
\sum_{j=1}^{d}\left\vert j\right\rangle _{s}\left\vert j\right\rangle
_{s^{\prime }}$ is a maximally correlated state of the $d$-dimensional open
system $s$ ($=ab$ in this paper) and an ancillary system $s^{\prime }.$
According to Eq. (\ref{map}), we have $h(t)=\partial _{t}|{\kappa }%
(t)|/|\kappa (t)|=\partial _{t}\ln |\kappa (t)|$ for $\partial _{t}|{\kappa }%
(t)|>0,$ otherwise $h(t)=0.$ Since the function $\ln |\kappa (t)|$ has the
same monotonicity as $|\kappa (t)|$, this divisibility based
non-Markovianity measure will yield the same result as that for the
information-flow based measure.

There is also an entanglement based non-Markovianity measure defined as \cite%
{RHP}
\begin{equation}
\mathcal{N}_{E}(\Lambda ):=\int_{\partial _{t}E>0}\partial _{t}{E}(\rho
^{ss^{\prime }}(t))dt
\end{equation}%
with $\rho ^{ss^{\prime }}(t)=(\Lambda _{t}\otimes \mathbf{I})\rho
^{ss^{\prime }}$ and $E(\cdot )$ a measure of entanglement. If we take $%
E(\cdot )$ as the negativity \cite{Negativity}, then in our model, $E(\rho
^{ss^{\prime }}(t))=\left\vert \kappa (t)\right\vert +1/2,$ and $\partial
_{t}E(\rho ^{ss^{\prime }}(t))=\partial _{t}\left\vert {\kappa }%
(t)\right\vert ,$ which will also yield the same negative non-Markovian
effect as that for the information-flow based measure.

We further consider the correlations based non-Markovianity measure \cite%
{LFS}
\begin{equation}
\mathcal{N}_{I}(\Lambda ):=\int_{\partial _{t}I>0}\partial _{t}I(\rho
^{ss^{\prime }}(t))dt,
\end{equation}%
where $\rho ^{ss^{\prime }}(t)=(\Lambda _{t}\otimes \mathbf{I})\rho
^{ss^{\prime }},$ $I(\rho ^{ss^{\prime }}(t))=S(\rho ^{s}(t))+S(\rho
^{s^{\prime }}(t))-S(\rho ^{ss^{\prime }}(t))$ is the quantum mutual
information, $\rho ^{s}(t)=\mathrm{tr}_{s^{\prime }}\rho ^{ss^{\prime }}(t)$%
, $\rho ^{s^{\prime }}(t)=\mathrm{tr}_{s}\rho ^{ss^{\prime }}(t)$, and $%
S(\rho ^{s}(t)):=-\mathrm{tr}\rho ^{s}(t)\log _{2}\rho ^{s}(t)$ is the von
Neumann entropy. After simple calculations, we have $I(\rho ^{ss^{\prime
}}(t))=4-H\left( \frac{1-\left\vert \kappa (t)\right\vert }{2}\right) ,$
where $H(x):=-x\log _{2}x-(1-x)\log _{2}(1-x).$ Since $0\leq \frac{%
1-\left\vert \kappa (t)\right\vert }{2}\leq \frac{1}{2},$ the monotonicity
of $I(\rho ^{ss^{\prime }}(t))$ is the same as $\left\vert \kappa
(t)\right\vert .$ In addition, $\partial _{t}I(\rho ^{ss^{\prime }}(t))=%
\frac{\partial _{t}\left\vert {\kappa }(t)\right\vert }{2}\log _{2}\frac{%
1+\left\vert \kappa (t)\right\vert }{1-\left\vert \kappa (t)\right\vert },$
which implies the same conclusion as that for the information-flow based
measure.

Obviously, $h(t),$ $E(\rho ^{ss^{\prime }}(t))$ and $I(\rho ^{ss^{\prime
}}(t))$ have the same monotonicity as $|\kappa (t)|$, and the above three
different non-Markovianity measures are all equivalent to the
information-flow measure for identifying non-Markovianity in our model. This
fact implies that the illustrated the negative effect of non-Markovianity is
a rather generic feature.

\section{Non-Markovian positive effect}

As an example of non-Markovian positive effect, we consider a two-qubit
system suffering from the quantum dynamics $\Lambda $ described by Eq. (\ref%
{DD}) with $f(\omega )$ a resonant Lorentz reservoir of frequency spectral
width $\Gamma $ and correlation time $\gamma _{0}^{-1}$ \cite{Book-Open}.

\begin{figure*}[tbp]
\centering
\includegraphics[width=10cm]{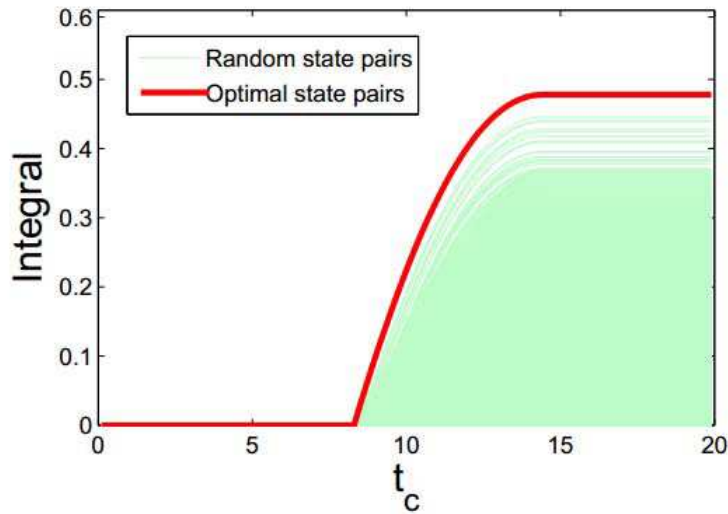}
\caption{(Color online) The integrals in the definition of non-Markovianity
versus control time $t_{c}$ of a two-qubit system with parameter $\Gamma /%
\protect\gamma _{0}=0.1$ as an example. The red and green curves represent
the non-Markovianity with state-pairs given by Eq. (\protect\ref{optimal})
and other (10000) randomly generated pairs, respectively. }
\label{fig:wide}
\end{figure*}

In Fig. 4, the integrals in the definition of $\mathcal{N}(\Lambda )$ (green
curves) of 10000 random state pairs are generated with $\Gamma /\gamma
_{0}=0.1$. The optimal state-pairs are still $\rho _{1,2}^{ab}=\left\vert
\zeta ^{\pm }\right\rangle \left\langle \zeta ^{\pm }\right\vert $ or $%
\left\vert \eta ^{\pm }\right\rangle \left\langle \eta ^{\pm }\right\vert $
[Eq. (\ref{optimal})], and we have
\begin{equation}
D(t)=|\chi (t)|\ \ \mathrm{with}\ \ \chi (t)=e^{-\frac{\Gamma t}{2}}\Big (%
\cos \frac{\varepsilon t}{2}+\frac{\Gamma }{\varepsilon }\sin \frac{%
\varepsilon t}{2}\Big ),
\end{equation}%
where $\varepsilon =\sqrt{|\Gamma ^{2}-2\gamma _{0}\Gamma |}.$

\begin{figure*}[tbp]
\centering
\includegraphics[width=10cm]{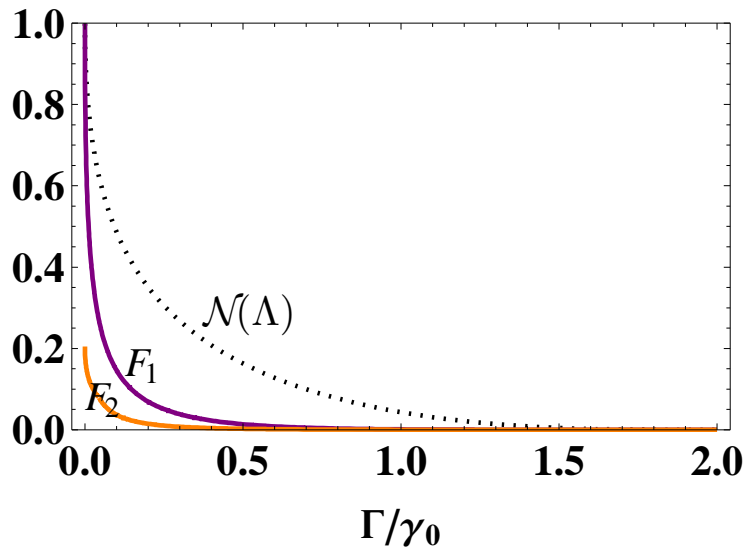}
\caption{(Color online) A case of non-Markovian positive effect: a two-qubit
system with qubit $A$ suffering from noise environment with a resonant
Lorentzian spectral distribution and qubit $B$ free from noise. The dotted
curve depicts the non-Markovianity with control time $t_{c}=2\protect\pi /%
\protect\varepsilon $ ($\protect\varepsilon =\protect\sqrt{|\Gamma ^{2}-2%
\protect\gamma _{0}\Gamma |}$) versus the parameter $\Gamma /\protect\gamma %
_{0}\in \lbrack 0,2]$ (non-Markovian regime). The purple (dark) and orange
(light) curves represent the RSP-fidelity $F_{1}$ and $F_{2}$ with initial
Bell state $(c_{1},c_{2},c_{3})=(1,-1,1)$ and a mixed entangled state $%
(c_{1},c_{2},c_{3})=(-0.5,0.4,0.8),$ respectively.}
\label{fig:wide}
\end{figure*}

For simplicity, we take $t_{c}=2\pi /{\varepsilon }$ as the control time,
then $\mathcal{N}(\Lambda )=|\chi (t_{c})|=e^{-\pi \Gamma /\varepsilon }$.
We still consider the RSP fidelity. If the initial state is the
Bell-diagonal state described by Eq. (\ref{BD}), then we have $F=F(C^{T}C)$
with the three eigenvalues $c_{1}^{2}\mathcal{N}^{2}(\Lambda )$, $c_{2}^{2}%
\mathcal{N}^{2}(\Lambda )$ and $c_{3}^{2}\mathcal{N}^{4}(\Lambda )$ of $%
C^{T}C.$ In Fig. 5, the fidelity with initial states ($c_{1},c_{2},c_{3}$)$%
=(1,-1,1)$ and $(-0.5,0.4,0.8)$ is depicted in comparison with the
non-Markovianity $\mathcal{N}(\Lambda )$ in the non-Markovian regime $\Gamma
/\gamma _{0}\in \lbrack 0,2]$ (smaller value corresponds to stronger
non-Markovian effect) \cite{Book-Open}. We see that non-Markovianity in this
example is positive: stronger non-Markovianity will induce higher RSP
fidelity.

\section{Conclusions}

In summary, we have revealed that non-Markovian effect in an experimentally
controllable photonic open system can have quite different virtues, stronger
non-Markovianity may cause lower performance of quantum tasks such as the
fidelity for remote state preparation. That is quite different from the
usual positive non-Markovian effect. Rigorously identifying the border
between the positive and negative effects of non-Markovian behavior is still
an open question, which is of great importance in open systems engineering.

This work was supported by NNSFC (No. 11204196 and No. 11074184) and SRFDPHE
(No. 20123201120004).

\end{document}